\documentclass[conference]{IEEEtran}
\usepackage{epsfig,makeidx,color}
\usepackage{algpseudocode}
\usepackage{url}
\usepackage{graphicx}
\usepackage{graphics}
\usepackage{latexsym}
\usepackage{showidx}
\usepackage{latexsym}
\usepackage{amssymb}
\usepackage{amsfonts}
\usepackage{amsmath}
\usepackage{graphicx}
\usepackage{amssymb}
\usepackage{amsfonts}
\usepackage{amsmath}
\ifCLASSINFOpdf
\else
\fi
\newtheorem{Theorem}{Theorem}

\newtheorem{Proposition}{Proposition}

\newtheorem{lem}{Lemma}

\newcommand{\rmv}[1]{}

\begin{document}
\title{Phase Precoded Compute-and-Forward with Partial Feedback}
\author{\authorblockN{Amin Sakzad, Emanuele Viterbo}\authorblockA{Dept. Elec. \& Comp. Sys.\\Monash University, Australia\\
{\tt amin.sakzad,emanuele.viterbo@monash.edu}}
\and
\authorblockN{Joseph Boutros,}\authorblockA{Dept. Electrical Engineering,\\
Texas A\&M University at Qatar\\
{\tt boutros@tamu.edu}}\and
\authorblockN{Yi Hong}\authorblockA{Dept. Elec. \& Comp. Sys.\\
Monash University, Australia\\
{\tt yi.hong@monash.edu}} }

\maketitle
\begin{abstract}
In this work, we propose phase precoding for the compute-and-forward (CoF) protocol. We derive the phase precoded computation rate and show that it is greater than the original computation rate of CoF protocol without precoder. To maximize the phase precoded computation rate, we need to `jointly' find the optimum phase precoding matrix and the corresponding network equation coefficients. This is a mixed integer programming problem where the optimum precoders should be obtained at the transmitters and the network equation coefficients have to be computed at the relays. To solve this problem, we introduce phase precoded CoF with partial feedback. It is a quantized precoding system where the relay jointly computes both a quasi-optimal precoder from a finite codebook and the corresponding network equations. The index of the obtained phase precoder within the codebook will then be fedback to the transmitters. A ``deep hole phase precoder'' is presented as an example of such a scheme. We further simulate our scheme with a lattice code carved out of the Gosset lattice and show that significant coding gains can be obtained in terms of equation error performance.
\end{abstract}

\begin{IEEEkeywords}
Compute-and-forward, lattice codes, phase precoding.
\end{IEEEkeywords}

\section{Introduction}
The rapid expansion of wireless networks and their application has promoted researchers to deal with more complex channel models including multi-terminal relay channels~\cite{Liew11}. In this framework, diversity techniques are used to combat channel fading. Different cooperative transmission protocols can be employed. In this paper, we focus on the recently proposed Compute-and-Forward (CoF) protocol~\cite{Nazer11} which maximizes the network throughput. This scheme uses algebraic structured codes to both harness the interference and remove the noise.

In CoF, the transmitters employ an identical lattice code and relays use the corresponding lattice decoder. For example, in a two-user case, suppose that ${\bf x}_1$ and ${\bf x}_2$ are the transmitted lattice codewords from the first and the second user, respectively. The received vector at the relay is $h_1{\bf x}_1+h_2{\bf x}_2+{\bf z}$ where ${\bf z}$ is the Guassian noise and the components of ${\bf h}=(h_1,~h_2)$ are the fading channel coefficients from the first and the second user to the relay, respectively. The task of the relay is to estimate an integer linear combination $a_1{\bf x}_1+a_2{\bf x}_2$ from the received vector. The estimated point $a_1{\bf x}_1+a_2{\bf x}_2$ is still a lattice vector because any integer linear combination of lattice points is lattice point. The quality of such an estimate and consequently the achievable computation rate is controlled by a non-zero coefficient $\alpha$. In particular, the parameter $\alpha$ and the integer vector ${\bf a}=(a_1,~a_2)$ are chosen so that $\alpha{\bf h} \approx {\bf a}$. This approximation comes with a penalty since the components of ${\bf a}$ are restricted to be integers only. In other words, the approximant space for $\alpha{\bf h}$ is the set of all integer vectors. This penalty is equivalent to the approximation of real vectors by rational ones and hence limits the computation rate in CoF protocol~\cite{Niesen}.

In this paper, we propose phase precoding for CoF protocol to increase the computation rate. We assume that the precoder for each transmitter is a complex scalar $e^{i\phi}$, for some $-\pi/4\leq\phi\leq\pi/4$, multiplying the lattice codeword. For example, in the two-user case, we send $e^{i\phi_1}{\bf x}_1$ and $e^{i\phi_2}{\bf x}_2$ instead of ${\bf x}_1$ and ${\bf x}_2$. The equivalent channel coefficient vector is ${\bf h}'=(e^{i\phi_1}h_1,~e^{i\phi_2}h_2)$. The parameters $\alpha'$ and ${\bf a}'$ have to be selected such that the quality of the new approximation $\alpha'{\bf h}'\approx {\bf a}'$ will be better than the original approximation $\alpha{\bf h} \approx {\bf a}$. More precisely, the precoders should be chosen so that the components of ${\bf h}'$ will be more aligned with Guassian integers. This alignment of ${\bf h}'$ and ${\bf a}'$ results in a higher computation rate which we call \emph{phase precoded computation rate}.

Our contributions are: ({\em i}) we introduce the concept of phase precoding for CoF protocol, ({\em ii}) we find the phase precoded computation rate and show it is greater than the original computation rate for CoF, ({\em iii}) we propose phase precoded CoF with partial feedback and as an example of this scheme, the deep hole phase precoder is presented, ({\em v}) we simulate our phase precoder scheme using lattice encoders and present numerical results.

{\bf Notation.} Boldface letters are used for vectors, and capital boldface letters for matrices. Superscripts $^T$ and $^H$ denote transposition and Hermitian transposition. $\mathbb{Z}$, $\mathbb{C}$, $\mathbb{R}$, and $\mathbb{Z}[i]$ denote the ring of rational integers, the field of complex numbers, the field of real numbers, and the ring of Gaussian integers, respectively. We let $|z|$ and $\mbox{arg}(z)$ denote the modulus and the phase of the complex number $z$, respectively. The Hermitian product of two row vectors ${\bf a}$ and ${\bf b}$ is denoted by $\langle{\bf a}, {\bf b}\rangle \triangleq {\bf a} {\bf b}^{H}$. The notation $\|{\bf v} \|$ stands for the Euclidean norm of the vector ${\bf v}$. Given a positive number $x$, we define $\log^+(x)\triangleq \max\{\log(x),0\}$. Finally, a $k\times k$ matrix ${\bf X}=\left({\bf x}_1^T|\cdots|{\bf x}_k^T \right)^T$ is formed by stacking the $k$-dimensional row vectors ${\bf x}_1,\ldots, {\bf x}_k$, and ${\bf I}_k$ denotes the $k\times k$ identity matrix.

\section{System Model}~\label{sec:back}
We recall the notion of lattice code which is essential throughout the paper. A $k$-dimensional {\em complex lattice} $\Lambda$ with {\em generator matrix}
${\bf G}\triangleq \left(\begin{array}{c|c|c|c}
{\bf g}_1^T&{\bf g}_2^T&\cdots &{\bf g}_k^T
\end{array}\right)^T,$
for ${\bf g}_j \in \mathbb{C}^n$ and $1\leq j\leq k$, is the set of points in $\mathbb{C}^n$
$$\Lambda=\{{\bf x}={\bf u}{\bf G}| {\bf u}\in \mathbb{Z}[i]^n\}.$$
If $n=k$, the lattice is called {\em full rank}. Around each lattice point ${\bf x}\in\Lambda$ is the {\em Voronoi region}
$$\nu({\bf x})=\left\{{\bf y}\in\mathbb{C}^n\colon\|{\bf y}-{\bf x}\|\leq\|{\bf y}-\boldsymbol{\lambda}\|,~\forall\boldsymbol{\lambda}\in\Lambda\right\}.$$

A subset $\Lambda'\subseteq\Lambda$ is called a {\em sublattice} if $\Lambda'$ is a lattice itself. Given a sublattice $\Lambda'$, we define the {\em lattice code} $\Lambda/\Lambda'$. This quotient includes a finite constellation of lattice points carved from the lattice $\Lambda$. A common choice~\cite{Conway83} for the sublattice $\Lambda'$ is $a\Lambda$ for some $a\in\mathbb{Z}[i]$. The shape of this constellation is determined by the Voronoi region of the lattice $\Lambda'$. For a vector ${\bf y}\in\mathbb{C}^n$, the {\em nearest-neighbor quantizer associated with $\Lambda$} is defined as
\begin{equation}~\label{Q}
Q_{\Lambda}({\bf y}) \triangleq \arg\!\min_{\boldsymbol{\lambda}\in\Lambda}\|{\bf y}-\boldsymbol{\lambda}\|.
\end{equation}
We also define the {\em modulo lattice operation} as
$${\bf y}\!\!\!\mod \Lambda\triangleq{\bf y}-Q_{\Lambda}({\bf y}).$$
\subsection{The compute-and-forward protocol}
\begin{figure}[htb]
\begin{center}
\includegraphics[width=7.5cm]{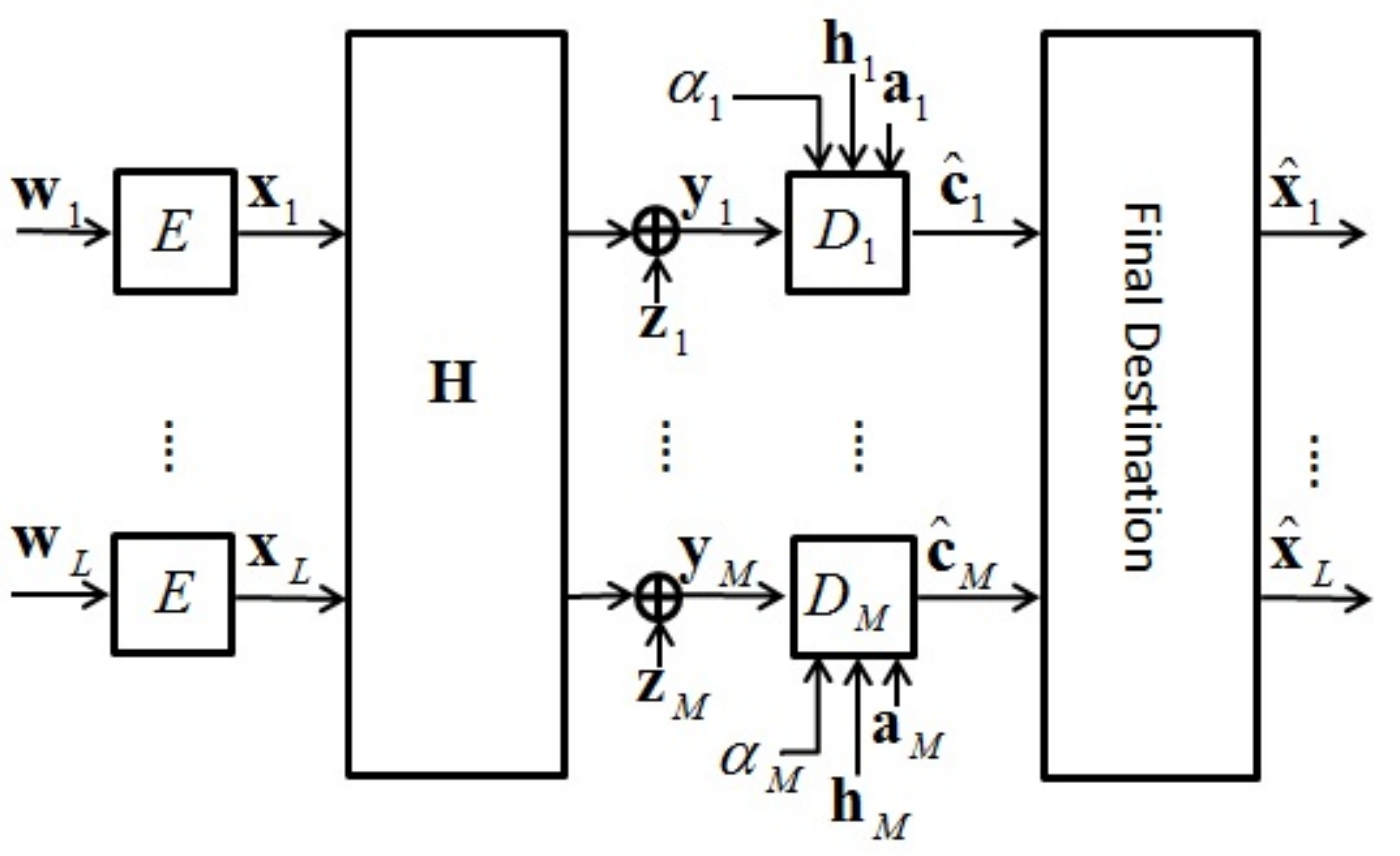}
\caption{\label{CF} The CoF protocol with $L$ transmitters and $M$ relays explained in \eqref{eq:sysmodel}.}
\end{center}
\end{figure}

Fig.~\ref{CF} illustrates a compute-and-forward (CoF) protocol~\cite{Nazer11} with $L$ transmitters and $M$ relay nodes. The $M$ relays compute estimates of $M$ linear equations of the transmitted information. These will be forwarded to the final destination, where they form a system of linear equations to recover the $L$ distinct messages. It is required that $M\geq L$, in order to be able to solve the system of $M$ linear equations with $L$ unknown variables.

In the CoF protocol, the $\ell$-th transmitter is equipped with an encoder $E\!:\mathbb{F}^{k}\rightarrow\Lambda/\Lambda'\subseteq\mathbb{C}^n$, where $\mathbb{F}$ is a finite field and $n$ is the codeword length. The encoder $E$ maps an information symbol vector ${\bf w}_\ell\in \mathbb{F}^{k}$ to a lattice codeword $E({\bf w}_\ell)={\bf x}_\ell\in\Lambda/\Lambda'$, for $1\leq\ell\leq L$. Each codeword is subject to the power constraint $\|{\bf x}_\ell\|^2\leq n\rho$. The $m$-th relay observes a noisy linear combination of the transmitted signals,
\begin{equation}~\label{eq:sysmodel}
{\bf y}_m=\sum_{\ell=1}^Lh_{m,\ell}{\bf x}_\ell+{\bf z}_m,
\end{equation}
where $h_{m,\ell}\in\mathbb{C}$, for $1\leq\ell\leq L$ and $1\leq m\leq M$, is the Rayleigh fading $\sim\mathcal{N}_{\mathbb{C}}(0,1)$ channel coefficient from $\ell$-th transmitter to the $m$-th relay and ${\bf z}_m$ is identically and independently distributed (i.i.d.) Gaussian complex noise $\mathcal{N}_{\mathbb{C}}(0,1)$.
The task of the $m$-th relay is to estimate a linear combination $\sum_{\ell=1}^La_{m,\ell}{\bf x}_\ell$ of the transmitted signals given an integer coefficient vector ${\bf a}_m\triangleq(a_{m,1},\ldots,a_{m,L})\in\mathbb{Z}[i]^L$ for all $1\leq m\leq M$. Due to the linear structure of lattices, the integer linear combinations are still in $\Lambda$ but not necessarily in the lattice code $\Lambda/\Lambda'$. At the $m$-th relay a detector
\begin{equation}~\label{eq:Decoder}
D_{m}:\mathbb{C}\times\mathbb{C}^L\times\mathbb{C}^n\times\mathbb{Z}[i]^L\rightarrow\Lambda/\Lambda',
\end{equation}
is employed to find an estimate $\hat{\bf c}_m$ of the codeword linear combination
\[
{\bf c}_m \triangleq \left(\sum_{\ell=1}^La_{m,\ell}{\bf x}_\ell\right)\!\!\!\mod{\Lambda'},
\]
which is a point in $\Lambda/\Lambda'$. The quality of this estimation is controlled by a non-zero complex $\alpha_m$. The $m$-th decoder at the relay first computes
\begin{eqnarray}
\!\!\!\!\alpha_m{\bf y}_m \!\!\!\!\!\!&=&\!\!\!\!\!\! \sum_{\ell=1}^L\alpha h_{m,\ell}{\bf x}_\ell+\alpha{\bf z}_m\label{eq:useful}\\
\!\!\!\!&=& \!\!\!\!\!\!\underbrace{\sum_{\ell=1}^La_{m,\ell}{\bf x}_\ell}_{\mbox{useful term}}+\underbrace{\sum_{\ell=1}^L\left(\alpha_m h_{m,\ell} - a_{m,\ell}\right){\bf x}_\ell+\alpha_m{\bf z}_m}_{\mbox{effective noise}},\nonumber
\end{eqnarray}
and then sets
$$\hat{\bf c}_m\triangleq D_{m}(\alpha_m,{\bf h}_m,{\bf y}_m,{\bf a}_m)= Q_{\Lambda}(\alpha_m{\bf y}_m) \!\!\!\mod{\Lambda'},$$
where ${\bf h}_m\triangleq(h_{m,1},\ldots,h_{m,L})\in\mathbb{C}^L$ and $Q_\Lambda$ and $D_m$ are defined in \eqref{Q} and \eqref{eq:Decoder}, respectively. The estimate $\hat{\bf c}_m$ of ${\bf c}_m$ will be sent through the network. At the final destination, a system of linear equations
$$\left\{\begin{array}{l}
\left(\sum_{\ell=1}^La_{1,\ell}{\bf x}_\ell\right) \!\!\!\mod{\Lambda'}= \hat{\bf c}_1,\\
\hspace{.5cm}\vdots\\
\left(\sum_{\ell=1}^La_{M,\ell}{\bf x}_\ell\right)\!\!\!\mod{\Lambda'} = \hat{\bf c}_M,
\end{array}
\right.$$
needs to be solved to find lattice codewords estimates $\hat{\bf x}_\ell$. Finally, the map $E^{-1}$ is used to produce the estimates $\hat{\bf w}_\ell$ of information symbol vectors ${\bf w}_\ell$, for $1\leq \ell\leq L$. In this framework, we declare an {\em equation error} at the $m$-th relay, if $\hat{\bf c}_m\neq{\bf c}_m$, for $1\leq m\leq M$. This refers to the event of decoding to an incorrect lattice codeword ${\bf c}_m$.

We recall from~\cite{Nazer11} and~\cite{Feng} some results about the computation rate for the $m$-th relay using CoF protocol:
\begin{Proposition}~\label{prop:computerate}
For complex-valued AWGN networks with a channel coefficient vector ${\bf h}_m$ and a coefficient vector ${\bf a}_m\in\mathbb{Z}[i]^L$, the following {\em computation rate} $\mathfrak{R}(\rho,{\bf h}_m,{\bf a}_m)$ is achievable:
\begin{equation}\label{ComputationRate}
\max_{0\neq\alpha_m\in\mathbb{C}} \log^+\left(\frac{\rho}{\rho\|\alpha_m{\bf h}_m-{\bf a}_m\|^2+|\alpha_m|^2}\right).
\end{equation}\hfill$\Box$
\end{Proposition}
From \eqref{eq:useful}, we note that the average energy of the effective noise is
\begin{equation}~\label{eq:q}
Q({\bf a}_m,\alpha_m)=\rho\|\alpha_m{\bf h}_m-{\bf a}_m\|^2+|\alpha_m|^2,
\end{equation}
affects the computation rate. The computation rate, given ${\bf a}_m$, provided in the above proposition is uniquely maximized by choosing $\alpha_m$ to be the minimum mean square estimator (MMSE) coefficient~\cite{Nazer11}
\begin{equation}~\label{alphammse}
\alpha_{\mbox{\tiny MMSE}}=\frac{\rho \langle{\bf h}_m,{\bf a}_m\rangle}{1+\rho\|{\bf h}_m\|^2}.
\end{equation}
Substituting $\alpha_{\mbox{\tiny MMSE}}$ of \eqref{alphammse} into $\mathfrak{R}(\rho,{\bf h}_m,{\bf a}_m)$ yields,~\cite{Feng}
\begin{equation}~\label{RateGram}
\mathfrak{R}(\rho,{\bf h}_m,{\bf a}_m)=\log^+\left(\frac{1}{{\bf a}_m{\bf M}{\bf a}_m^H}\right),
\end{equation}
where ${\bf M}$ is
\begin{equation}~\label{Gram}
{\bf M}={\bf I}_L-\frac{\rho}{1+\rho\|{\bf h}_m\|^2}{\bf h}_m^H{\bf h}_m.
\end{equation}
\section{Phase Precoder for Compute-and-Forward}~\label{sec:PP}
\begin{figure}[htb]
\begin{center}
\includegraphics[width=7.5cm]{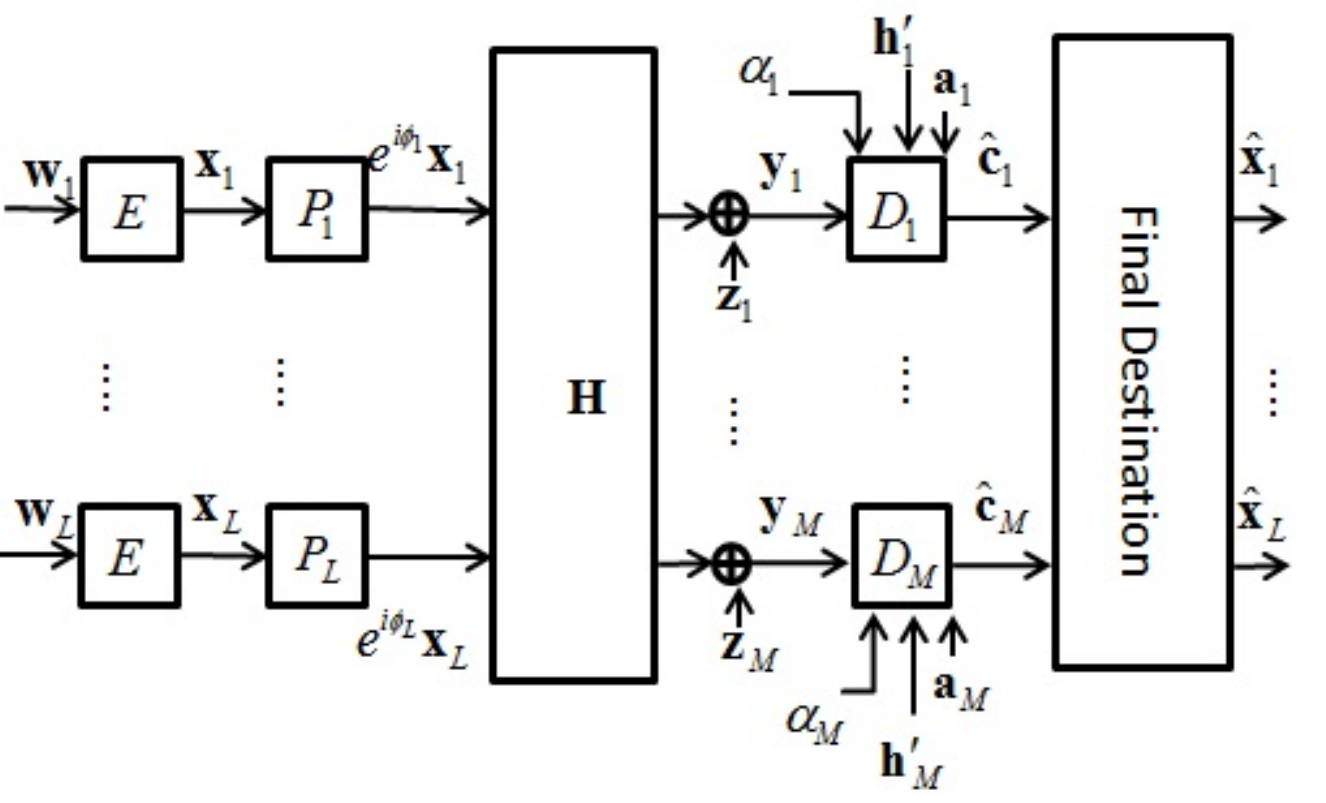}
\caption{\label{CFPP} Phase precoded CoF with $L$ transmitters and $M$ relay nodes.}
\end{center}
\end{figure}

Fig.~\ref{CFPP} illustrates a network with $L$ transmitters equipped with phase precoders (PP) and $M$ relays each employing CoF strategy. After the encoder $E$,
a lattice codeword ${\bf x}_\ell \in\Lambda/\Lambda'$, is generated at the $\ell$-th transmitter. We consider a block fading channel model, {\em i.e.} the channel coefficients ${\bf h}_m$ remain unchanged for a time frame of length $t\gg n$. These channel gains vary independently from one frame to the next. A frame header is used for the training phase, where we apply a phase precoding function $P_\ell\colon\mathbb{C}^n\rightarrow\mathbb{C}^n$, which maps ${\bf x}_\ell$ to $P_\ell({\bf x}_\ell)\triangleq e^{i \phi_{\ell}}{\bf x}_\ell$, for $\phi_\ell\in[-\pi/4,\pi/4]$ and $1\leq \ell\leq L$. Due to the symmetry of the complex plane, the problem of choosing the optimum network equation coefficients for CoF protocol can be reduced to the vectors ${\bf a}_m$ with components $a_{m,\ell}$ satisfying $\arg(a_{m,\ell})\in[-\pi/4,\pi/4]$. Thus, the phases for precoding can also be restricted to $e^{i \phi_{\ell}}$ with $\phi_\ell\in[-\pi/4,\pi/4]$. Using this approach, the phase precoded codeword $e^{i\phi_\ell}{\bf x}_\ell$ continues to satisfy the power constraint $\|e^{i\phi_\ell}{\bf x}_\ell\|^2\leq n\rho$, for $1\leq \ell \leq L$. Thus, the $m$-th relay receives
\begin{eqnarray}
{\bf y}_m &=& \sum_{\ell=1}^L h_{m,\ell}e^{i \phi_{\ell}}{\bf x}_\ell+{\bf z}_m.\label{eq:Phibetweenhandx}
\end{eqnarray}
We let $h'_{m,\ell}=h_{m,1}e^{i \phi_{\ell}}$, for $1\leq \ell \leq L$ and
$${\bf h}'_m\triangleq(h'_{m,1},\ldots,h'_{m,L})={\bf h}_m\Phi,$$
where
\begin{equation}\label{Phi}
{\Phi}\triangleq\mbox{diag}\left(e^{i \phi_{1}},\ldots,e^{i \phi_{L}}\right).
\end{equation}
As a result of considering the matrix $\Phi$ as part of ${\bf h}'_m$, the $m$-th relay recovers an integer linear combination $\sum_{\ell=1}^La_{m,\ell}{\bf x}_\ell$ of the transmitted codewords. Therefore, it first computes:
\begin{eqnarray}
{\bf y}'_m\!\!\!\!\!\!&=&\!\!\!\!\!\! \alpha_m{\bf y}_m = \sum_{\ell=1}^L\alpha_m h'_{m,\ell}{\bf x}_\ell+\alpha_m{\bf z}_m\label{eq:PPeffectivenoise}\\
&=&\!\!\!\!\!\! \sum_{\ell=1}^La_{m,\ell}{\bf x}_\ell+\underbrace{\sum_{\ell=1}^L\left(\alpha_m h'_{m,\ell}- a_{m,\ell}\right){\bf x}_\ell+\alpha_m{\bf z}_m.}_{\mbox{PP effective noise}} \nonumber
\end{eqnarray}
The $m$-th decoder $D_m$ will operate similarly to the CoF protocol except that it assumes ${\bf h}'_m$ rather than ${\bf h}_m$.
The {\em phase precoded computation rate} $\mathfrak{R}'(\rho,{\bf h}_m,\Phi,{\bf a}_m)$ for the $m$-th relay is defined as
\begin{equation}~\label{rateprecoder}
\max_{\alpha_m\in\mathbb{C}\setminus\{0\}}\log^+\left(\frac{\rho}{\rho\|\alpha_m{\bf h}'_m-{\bf a}_m\|^2+|\alpha_m|^2}\right).
\end{equation}
Based on \eqref{eq:PPeffectivenoise}, the average energy of the PP effective noise is
\begin{equation}~\label{firstminimization}
Q'({\Phi},{\bf a}_m,\alpha_m)=\rho\|\alpha_m{\bf h}'_m-{\bf a}_m\|^2+|\alpha_m|^2,
\end{equation}
which appears in the denominator of \eqref{rateprecoder}. Therefore, the $m$-th relay should calculate the best non-zero equalizer $\alpha_m\in\mathbb{C}$ and a non-zero network equation coefficient vector ${\bf a}_m\in\mathbb{Z}[i]^L$, to maximize \eqref{rateprecoder} or equivalently minimize \eqref{firstminimization}.
\subsection{Maximizing Phase Precoded Computation Rate}
There are three parameters $\alpha_m\in\mathbb{C}$, ${\bf a}_m\in\mathbb{Z}[i]^L$ and $\Phi=\mbox{diag}\left(e^{i \phi_{1}},\ldots,e^{i \phi_{L}}\right)$, where $\phi_\ell\in[-\pi/4,\pi/4]$, to be optimized. The selection procedure is based on two steps: ({\em i}) we suppose that ${\bf a}_m$ and $\Phi$ are fixed and find the optimum $\alpha_m$, then we substitute this optimum $\alpha_m$ into \eqref{rateprecoder}, ({\em ii}) we suppose ${\bf a}_m$ is given and find the best phases to maximize \eqref{rateprecoder}.

Replacing ${\bf h}_m$ by ${\bf h}'_m$ in Proposition~\ref{prop:computerate} and \eqref{alphammse}, given ${\bf a}_m\in\mathbb{Z}[i]^L$ and $\Phi$ as in \eqref{Phi} with $\phi_\ell\in[-\pi/4,\pi/4]$, the optimum $\alpha_{\tiny\mbox{opt}}'\in\mathbb{C}$ to minimize \eqref{firstminimization} is
\begin{equation}~\label{alphammseprime}
\alpha_{\tiny\mbox{opt}}'=\frac{\rho\left\langle{\bf h}'_m,{\bf a}_m\right\rangle}{1+\rho\|{\bf h}'_m\|^2}.
\end{equation}
Substituting $\alpha_{\tiny\mbox{opt}}'$ into \eqref{rateprecoder} yields
\begin{eqnarray}~\label{rateprecoderafter}
\!\!\!\!\!\!\!\!\!\!\!\!\!\!\!\!\!\!\!\!\!\!\!\!\!\!\!\!\!\!\!\!\!\!\!\mathfrak{R}'(\rho,{\bf h}_m,\Phi,{\bf a}_m)\!\!\!&=&\!\!\!\log^+\left(\frac{1}{{\bf a}_m\Phi^H{\bf M}\Phi{\bf a}_m^H}\right)\label{eq:aphiMphia}\\
\!\!\!\!\!\!\!\!\!\!\!\!\!\!\!\!\!\!\!\!\!\!\!\!\!\!\!\!\!\!\!\!\!\!\!&=&\!\!\!\log^+\left(1+\rho\|{\bf h}'_m\|^2\right)\nonumber
\end{eqnarray}
\begin{equation}\label{eq:aphiMphia1}
-\log^+\left(\|{\bf a}_m\|^2+\rho\left(\|{\bf h}'_m\|^2\|{\bf a}_m\|^2 -\left|\langle{\bf h}'_m,{\bf a}_m\rangle\right|^2\right)\right)\!\!,
\end{equation}
where ${\bf M}$ is given in \eqref{Gram} and ${\bf h}'_m={\bf h}_m\Phi$.
\begin{lem}\label{lem:fixeda}
Given the network equation coefficients
$${\bf a}_m=(a_{m,1},\ldots,a_{m,L})=(\beta_1e^{i\psi_1},\ldots, \beta_Le^{i\psi_L})\in\mathbb{Z}[i]^L,$$
and the channel coefficient
$${\bf h}_m=(h_{m,1},\ldots,h_{m,L})=(\eta_1e^{i\theta_1},\ldots, \eta_Le^{i\theta_L})\in\mathbb{C}^L,$$
the optimal phases to maximizing the phase precoded computation rate \eqref{rateprecoder} are $\phi_\ell=\psi_\ell-\theta_\ell$, for $1\leq \ell\leq L$.
\end{lem}
\begin{IEEEproof}
We prove this lemma for $L=2$. The proof for $L>2$ is similar to this case and we omit it for the sake of brevity. We find
$${\bf h}'_m\!=\!\left(h_{m,1}e^{i\phi_1},h_{m,2}e^{i\phi_2}\right)\!=\!\left(\eta_1e^{i(\theta_1+\phi_1)},\eta_2e^{i(\theta_2+\phi_2)}\right).$$
Based on \eqref{eq:aphiMphia1}, we have to maximize $\left|\langle{\bf h}'_m,{\bf a}_m\rangle \right|^2$ to achieve the highest computation rate. We have that
\vspace{.3cm}
$$\hspace{-7cm}\left|\langle{\bf h}'_m,{\bf a}_m\rangle \right|^2$$
\vspace{-.7cm}\begin{eqnarray}~\label{UBlll}
\!\!\!\!\!\!&=&\!\!\!\!\left|h'_{m,1}a_{m,1}^H+h'_{m,2}a_{m,2}^H\right|^2\nonumber\\
\!\!\!\!\!\!&=&\!\!\!\!\left|\eta_1e^{i(\theta_1+\phi_1)}\beta_1e^{i(-\psi_1)}+\eta_2e^{i(\theta_2+\phi_2)}\beta_2e^{i(-\psi_2)}\right|^2\nonumber\\
\!\!\!\!\!\!&=&\!\!\!\!\left|\left(\eta_1\beta_1\cos(\theta_1+\phi_1-\psi_1)+\eta_2\beta_2\cos(\theta_2+\phi_2-\psi_2)\right.\right.\nonumber\\
\!\!\!\!\!\!&+&\!\!\!\!\left.\left.i(\eta_1\beta_1\sin(\theta_1+\phi_1-\psi_1)+\eta_2\beta_2\sin(\theta_2+\phi_2-\psi_2)\right)\right|^2\nonumber\\
\!\!\!\!\!\!&=&\!\!\!\!\eta_1^2\beta_1^2\!+\!2\eta_1\eta_2\beta_1\beta_2\cos(\theta_1+\phi_1-\psi_1-(\theta_2+\phi_2-\psi_2))\nonumber\\
\!\!\!\!\!\!&+&\!\!\!\!\eta_2^2\beta_2^2\nonumber,
\end{eqnarray}
which means that in order to maximize $\left|\langle{\bf h}'_m,{\bf a}_m\rangle \right|^2$, the phases $\phi_1$ and $\phi_2$ have to satisfy
\begin{equation}~\label{eq:just}
\theta_1+\phi_1-\psi_1=\theta_2+\phi_2-\psi_2.
\end{equation}
Thus, we get $\phi_1^{\tiny\mbox{opt}}=\psi_1-\theta_1$ for the first transmitter and $\phi_2^{\tiny\mbox{opt}}=\psi_2-\theta_2$ for the second transmitter.
\end{IEEEproof}
\begin{Theorem}~\label{th:rateprecoder}
Given the channel coefficients ${\bf h}_m\in\mathbb{C}^L$, signal-to-noise ratio $\rho$, and the network equation coefficient vector ${\bf a}_m\in\mathbb{Z}[i]^L$, the phase precoded computation rate
$$\hspace{-2.5cm}\mathfrak{R}'(\rho,{\bf h}_m,\Phi^{\tiny\mbox{opt}},{\bf a}_m)=\log\left(1+\rho\|{\bf h}_m\|^2\right)-$$
\begin{equation}\label{eq:final}
\log\left(\!\|{\bf a}_m\|^2\!+\!\rho\left(\|{\bf h}_m\|^2\|{\bf a}_m\|^2\!-\!\left(\sum_{\ell=1}^L|h_{m,\ell}||a_{m,\ell}|\right)^2\right)\right).
\end{equation}
is greater than $\mathfrak{R}(\rho,{\bf h}_m,{\bf a}_m)$ where
\begin{equation}~\label{eq:phitopopt}
\Phi^{\tiny\mbox{opt}}\triangleq\mbox{diag}\left(e^{i\phi^{\tiny\mbox{opt}}_1},\ldots,e^{i\phi^{\tiny\mbox{opt}}_L}\right).
\end{equation}
\end{Theorem}
\begin{IEEEproof}
The computation rate without phase precoder is
$$\!\!\!\!\!\mathfrak{R}(\rho,{\bf h}_m,{\bf a}_m)=\log^+\left(1+\rho\|{\bf h}_m\|^2\right)$$
$$-\log^+\left(\|{\bf a}_m\|^2+\rho\left(\|{\bf h}_m\|^2\|{\bf a}_m\|^2 -\left|\langle{\bf h}_m,{\bf a}_m\rangle\right|^2\right)\right).$$
If we use phases $\phi_\ell=\theta_\ell-\psi_\ell$, for $1\leq \ell\leq L$, then we get
$$\left|\langle{\bf h}'_m,{\bf a}_m\rangle \right|^2= \eta_1^2\beta_1^2+2\eta_1\eta_2\beta_1\beta_2+\eta_2^2\beta_2^2.$$
On the other hand,
\vspace{.3cm}
$$\hspace{-7cm}\left|\langle{\bf h}_m,{\bf a}_m\rangle \right|^2$$
\vspace{-.7cm}\begin{eqnarray}~\label{UBlll1}
&=&\!\!\!\!\! \left|h_{m,1}a_{m,1}^H+h_{m,2}a_{m,2}^H\right|^2\nonumber\\
&=&\!\!\!\!\! \left|\eta_1e^{i(\theta_1)}\beta_1e^{i(-\psi_1)}+\eta_2e^{i(\theta_2)}\beta_2e^{i(-\psi_2)}\right|^2\nonumber\\
&=&\!\!\!\!\! \left|\left(\eta_1\beta_1\cos(\theta_1-\psi_1)+\eta_2\beta_2\cos(\theta_2-\psi_2) \right.\right.\nonumber\\
&+&\!\!\!\!\! \left.\left.i(\eta_1\beta_1\sin(\theta_1-\psi_1)+\eta_2\beta_2\sin(\theta_2-\psi_2))\right)\right|^2\nonumber\\
&=&\!\!\!\!\! \eta_1^2\beta_1^2+2\eta_1\eta_2\beta_1\beta_2\cos(\theta_1-\psi_1+\theta_2-\psi_2)+\eta_2^2\beta_2^2\nonumber.
\end{eqnarray}
It is clear that $\left|\langle{\bf h}'_m,{\bf a}_m\rangle \right|^2\geq \left|\langle{\bf h}_m,{\bf a}_m\rangle \right|^2$, which implies
that $\mathfrak{R}'(\rho,{\bf h}_m,\Phi,{\bf a}_m)\geq \mathfrak{R}(\rho,{\bf h}_m,{\bf a}_m)$.
\end{IEEEproof}
To maximize the phase precoded computation rate, the optimum phase precoder matrix and the corresponding network equation coefficients should be computed jointly. This is a mixed integer programming problem because the entries of the phase precoding matrix $\Phi$ are complex numbers and the components of ${\bf a}_m$ are Gaussian integers. In addition, the phase precoders need to be optimized at the transmitters and the integer coefficients have to be computed at the relay. Recalling Lemma~\ref{lem:fixeda}, for a given ${\bf a}_m$, the optimum $\Phi^{\tiny\mbox{opt}}$ can be derived as \eqref{eq:phitopopt}. However, this needs the knowledge of ${\bf a}_m$ at the transmitters. On the other hand, using \eqref{eq:aphiMphia} for a fixed $\Phi$, a method of finding the optimum ${\bf a}_m$ is to consider ${\bf M}' = \Phi^H{\bf M}\Phi$ and employ one of the approaches presented in~\cite{Feng,sakzad12}. This means that the optimum ${\bf a}_m$ can only be computed at the relays when the optimum $\Phi$ was known at the transmitters. Hence, a systematic approach of maximizing the phase precoded computation rate by optimizing both the Gaussian integer vector ${\bf a}_m$ and $\Phi$ `jointly' is not available. We then introduce partial feedback phase precoders for CoF. This is a quantized precoding system where a quasi-optimal precoder is chosen from a finite codebook of phases at the relay. The index of the best precoder is transferred from the relay to the transmitters over a feedback link. Criteria are provided for selecting the
optimal precoding matrix based on the phase precoded computation rate.
\subsection{Phase Precoders with partial feedback}
In a phase precoded CoF with $L$ users and one relay, we suppose that only a finite set of phases
$$\mathcal{S}=\left\{\overline{\phi}_1=0,\ldots,\overline{\phi}_s,\ldots,\overline{\phi}_{|\mathcal{S}|}\right\}\subseteq[-\pi/4,\pi/4]$$
is available at each transmitter. This corresponds to a finite codebook of phase precoders
$$\mathcal{C}=\left\{\Phi_c=\mbox{diag}\left(e^{i\overline{\phi}_{s_1}},\ldots,e^{i\overline{\phi}_{s_L}}\right)\colon \overline{\phi}_{s_\ell}\in\mathcal{S},~1\leq \ell \leq L\right\}$$
with $|\mathcal{C}|=|\mathcal{S}|^L$. Using a header with $|\mathcal{C}|$ pilot symbols the relay can select the best precoder from the codebook by computing
\begin{equation}~\label{eq:minQprime}
{\Phi}_{\tiny\mbox{opt}}=\arg\!\max_{\hspace{-.6cm}\Phi\in\mathcal{C}}\max_{{\bf a}_m\in\mathbb{Z}[i]^L}\mathfrak{R}'(\rho,{\bf h}_m,\Phi, {\bf a}_m).
\end{equation}
Let the selected precoder be
$${\Phi}_{\tiny\mbox{opt}}=\mbox{diag}\left(e^{i\phi'_{s_1}},\ldots,e^{i\phi'_{s_\ell}},\ldots,e^{i\phi'_{s_{L}}}\right),$$
where $\phi'_{s_\ell}\in\mathcal{S}$ for $1\leq s_\ell \leq |\mathcal{S}|$, then the relay feedbacks the index $s_\ell$ instead of $\phi'_{s_\ell}$ to the $\ell$-th transmitter, for $1 \leq \ell \leq L$. Therefore, we need at most $\log_2\left(|\mathcal{S}|\right)$ of feedback to be sent to each transmitter. Note that the codebook includes the non-precoded case ${\bf I}_L=\mbox{diag}\left(e^{i\overline{\phi}_{1}},\ldots,e^{i\overline{\phi}_{1}}\right)$, then we have:
\begin{eqnarray*}
\max_{{\bf a}_m\in\mathbb{Z}[i]^L}\mathfrak{R}'(\rho,{\bf h}_m,{\Phi}_{\tiny\mbox{opt}}, {\bf a}_m)&\geq&\max_{{\bf a}_m\in\mathbb{Z}[i]^L}\mathfrak{R}'(\rho,{\bf h}_m,{\bf I}_L, {\bf a}_m)\\
&=&\max_{{\bf a}_m\in\mathbb{Z}[i]^L}\mathfrak{R}(\rho,{\bf h}_m,{\bf a}_m).
\end{eqnarray*}
The above inequality guarantees that using this scheme we can increase the phase precoded computation rate in comparison with the original computation rate. We next provide an example of a phase precoder with partial feedback.
\subsection{Deep hole phase precoders}
\begin{figure}[htb]
\begin{center}
\includegraphics[width=3.7cm]{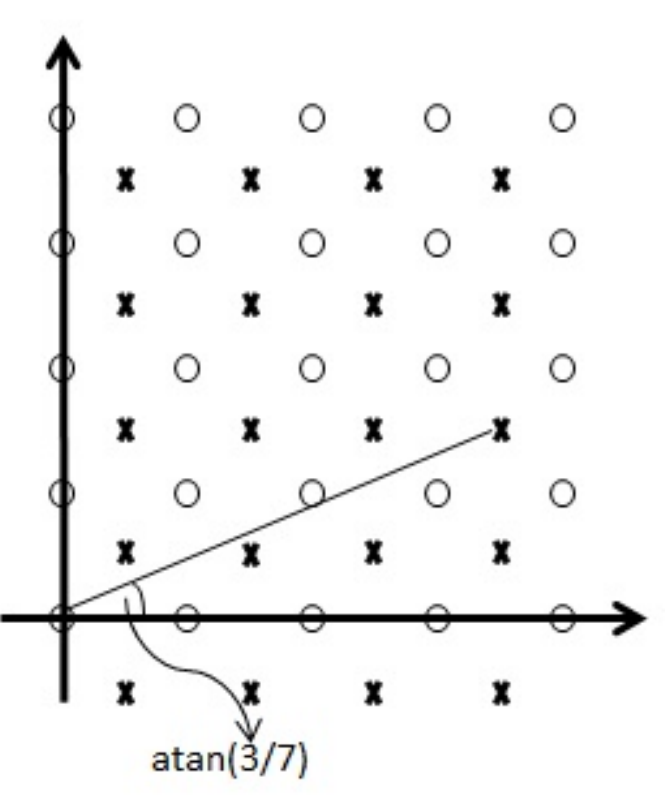}
\caption{\label{fig:DH} The lattice $\mathbb{Z}^2\simeq\mathbb{Z}[i]$ (empty circles) and its deep holes (solid circles).}
\end{center}
\end{figure}

A deep hole of an $n$-dimensional lattice $\Lambda$ is a point ${\bf x}$ whose distance
$\delta_2\left({\bf x},\Lambda\right)\triangleq\inf_{\boldsymbol{\lambda}\in\Lambda}\left\{\|{\bf x}-\boldsymbol{\lambda}\|\right\}$
is a global maximum. For example, the deep holes of $\mathbb{Z}^2$ are shown by crosses in Fig.~\ref{fig:DH}. In fact, a point $(o_1/2,o_2/2)$ for odd integers $o_1,o_2$ is a deep hole in $\mathbb{Z}^2$. We only consider odd integers $o_1$ and $o_2$ satisfying $o_2\leq o_1$. The corresponding phase of $(o_1/2,o_2/2)$ is $\mbox{atan}(o_2/o_1)$. For deep hole phase precoders, we use a finite number of deep hole phases of the lattice $\mathbb{Z}^2$ as $\mathcal{S}$.

\section{Simulation Results}~\label{sec:simulations}
In our simulations, we set $L=2$, $M=1$, $\Lambda=\mathcal{E}_8$~\cite{Conway83}, the densest lattice packing of dimension $8$, and $\Lambda'=a\mathcal{E}_8$, for $a=4$. Since our scheme works over complex numbers, the complex version of $\mathcal{E}_8$ can be identified~\cite{Conway83}. We use deep hole phase precoder with
$$\mathcal{S}=\left\{0,\pm\mbox{atan}(1/3),\pm\mbox{atan}(1/5),\pm\mbox{atan}(3/5),\mbox{atan}(1)\right\}.$$
and hence $|\mathcal{S}|=8$. The relay then feedback $\log_2(8)=3$ bits to each user providing the best phase to be used.

\begin{figure}[htb]
\begin{center}
\includegraphics[width=7cm]{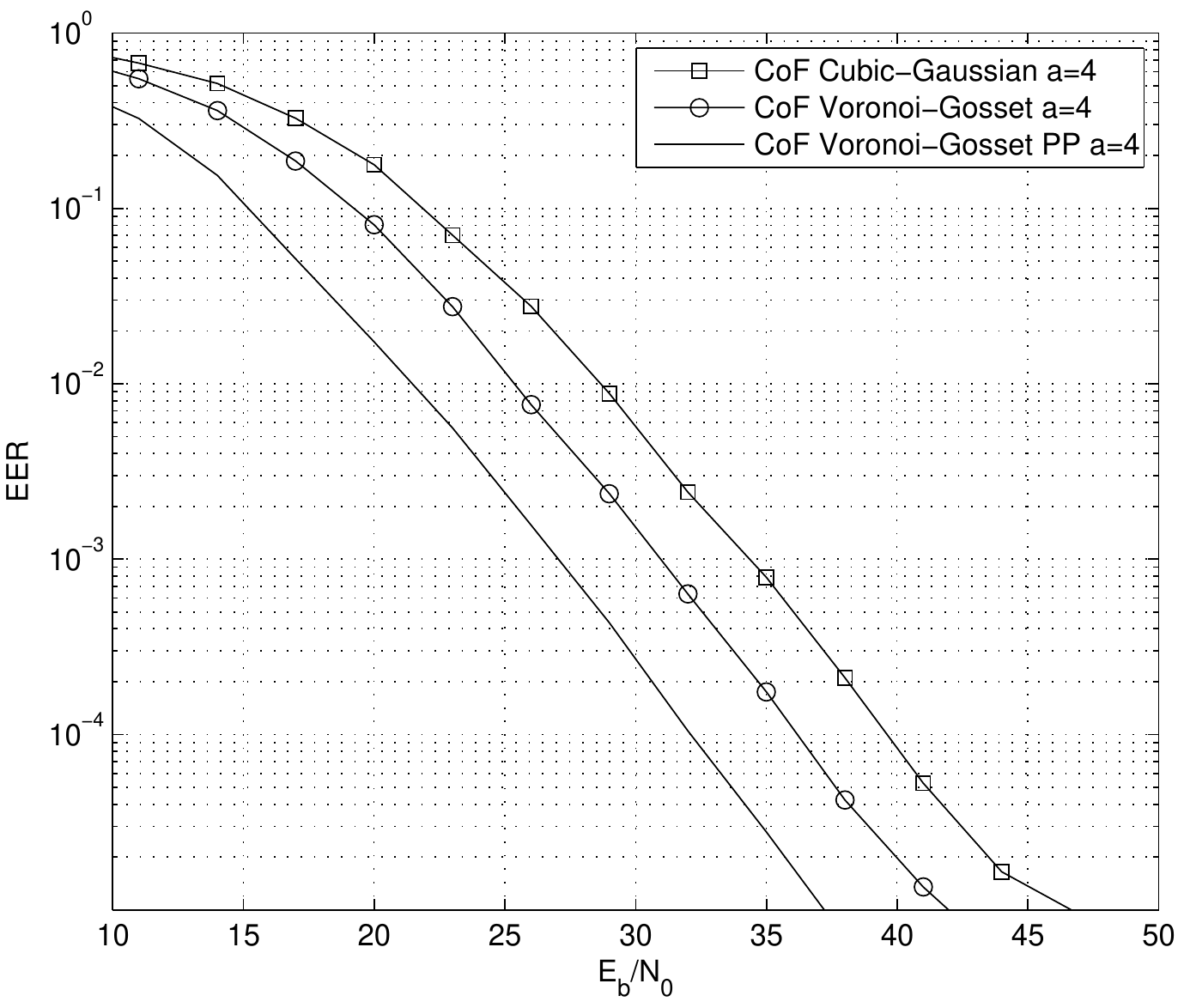}
\caption{~\label{fig:CoFE8P2m24}EER for $2$ users and $1$ relay node.}
\end{center}
\end{figure}

Fig.~\ref{fig:CoFE8P2m24} shows equation error rate (EER) for different lattice encoders including the cubic shaped Gaussian lattice $\mathbb{Z}[i]^4$ and Voronoi constellation carved from Gosset lattice $\mathcal{E}_8$ and its sublattices $a\mathcal{E}_8$ for $a=4$. This corresponds to rate $2$ bits per channel uses. To find equation coefficients, we have used a generalized version of QES presented in~\cite{sakzad12}.
Using Voronoi lattice constellation $\mathcal{E}_8/4\mathcal{E}_8$ for CoF protocol, we achieve $3.4$dB coding gain at EER of $10^{-4}$ in comparison with CoF over uncoded Cubic-Gaussian lattice $\mathbb{Z}[i]^L$. An extra $4$dB coding gain has also been obtained at EER of $10^{-4}$ using phase precoder over a CoF protocol equipped with $\mathcal{E}_8/4\mathcal{E}_8$ Voronoi lattice encoder.
\section{Conclusion}~\label{sec:conclusion}
A phase precoder scheme has been introduced for CoF protocol in physical layer network coding. The phase precoded computation rate has been derived. It has been shown that the proposed scheme achieve greater rate than that in~\cite{Nazer11}. Since the optimum phases and the optimum network equation coefficients to maximize the rate can not be captured easily, we suggested phase precoded CoF with partial feedback. Simulations were presented to show the effectiveness of the deep hole phase precoded CoF with partial feedback.

Investigating other aspects of phase precoded CoF protocol such as the degrees-of-freedom is also of interest. In addition, finding the optimum set $\mathcal{S}$ which maximizes the phase precoded achievable rate is the subject of future research studies.


\begin{thebibliography}{99}
\bibitem{Liew11} S.-C.~Liew, S.~Zhang, and L.~Lu,
``Physical-layer network coding: Tutorial, survey, and beyond,''
{\em Phys. Commun., 2011}. Available online at: http://arxiv.org/abs/1105.4261.

\bibitem{Nazer11} B.~Nazer and M.~Gastpar,
``Compute-and-Forward: Harnessing interference through structured codes,''
{\em IEEE Trans. on Inform. Theory,} vol.~57, pp.~6463--6486, 2011.

\bibitem{Niesen} U.~Niesen and P.~Whiting,
``The Degrees of Freedom for Compute-and-Forward,''
{\em IEEE Trans. on Inform. Theory,} vol.~58, pp.~5214--5232, 2012.


\bibitem{Feng} C.~Feng, D.~Silva, and F.R.~Kschichang,
``An algebraic approach to physical-layer network coding,''
{\em IEEE Trans. on Inform. Theory,} vol.,~59, pp.~7576--7596, 2013.

\bibitem{caire} S-N.~Hong, G.~Caire
``Compute-and-Forward Strategies for Cooperative Distributed Antenna Systems,''
{\em IEEE Trans. on Inform. Theory,} vol.~59, pp.~5227--5243, 2013.

\bibitem{Conway83}
J.~Conway and N.~Sloane, ``A fast encoding method for lattice codes and quantizers,''
{\em IEEE Trans. on Inform. Theory,} vol.~29, pp.~820-–824, 1983.

\bibitem{sakzad12} A.~Sakzad, E.~Viterbo, Y.~Hong, and J.J.~Boutros,
``On the ergodic rate for compute-and-forward,''
in {\em Proc. of International Symposium on Network Coding (NetCod 2012), MIT University, Boston, MA, USA,} pp.~131--136, 2012.


\end{thebibliography}
\end{document}